\documentclass[final,5p,times,twocolumn]{elsarticle}

\usepackage{amssymb}
\usepackage{amsmath}
\usepackage{hyperref}
\usepackage[normalem]{ulem}
\usepackage{color}

\newcommand{\changed}[1]{{#1}}

\journal{Fusion Engineering and Design}

\begin{document}

\begin{frontmatter}



\title{Efficient calculation of magnetic fields from ferromagnetic materials near strong electromagnets, and application to stellarator coil optimization}  


% \author{Matt Landreman, Humberto Torreblanca, and Antoine Cerfon} %% Author name
\author[UMD,T1E]{Matt Landreman}
\author[T1E]{Humberto Torreblanca}
\author[T1E]{Antoine Cerfon}

\affiliation[UMD]{organization={University of Maryland},%Department and Organization
            city={College Park},
            state={MD},
            postcode={20742}, 
            country={USA}}
\affiliation[T1E]{organization={Type One Energy Group, Inc.},%Department and Organization
            city={Knoxville},
            state={TN},
            postcode={37931}, 
            country={USA}}

%% Abstract
\begin{abstract}
In fusion reactor design, steels under consideration for the blanket are ferromagnetic, so the steel's effect on the plasma physics must be examined.
For efficient calculation of these fields, we can exploit the fact that the ferromagnetic material gives a small perturbation relative to the fields from the electromagnetic coils and plasma.
Moreover the magnetization is saturated due to the strong fields in typical fusion systems.
These approximations significantly reduce the nonlinearity of the problem, so the magnetic materials can be described by an array of point dipoles of known magnitude, oriented in the direction of the coil and plasma field.
The approach is verified by comparison to finite-element calculations with commercial software and shown to be accurate.
As no linear or nonlinear solve is required, only evaluation of Biot-Savart-type integrals, the method here is significantly simpler to implement than other methods, and extremely fast.
The method is compatible with arbitrary CAD geometry, and also allows rapid computation of the magnetic forces.
We demonstrate adding the ferromagnetic effects to free-boundary magnetohydrodynamic equilibrium calculations, assessing the effect on plasma physics properties such as confinement and stability.
Moreover, it is straightforward to differentiate through the model to get the derivative of the field with respect to the electromagnet parameters.
We thereby demonstrate gradient-based coil optimization for a quasi-isodynamic stellarator in which the field contribution from a ferromagnetic blanket is included.
Even a significant steel volume is found to have little impact on the plasma physics properties, with the main effects being a slight destabilization of ballooning modes and a radial shift of the edge islands due to decrease in rotational transform.
Both of these issues are corrected by the minor reoptimization of the coil shapes to account for the field from the steel.

\end{abstract}


% \begin{highlights}
% \end{highlights}

%% Keywords

\end{frontmatter}



%%%%%%%%%%%%%%%%%%%%%%%%%%%%%%%%%%%%%%%%%%%%%%%%%%%%%
%%%%%%%%%%%%%%%%%%%%%%%%%%%%%%%%%%%%%%%%%%%%%%%%%%%%%

\section{Introduction}
\label{sec:intro}

In the blanket of a fusion reactor, the leading candidates for structural materials are steels that are ferromagnetic \cite{tanigawa2017development,mergia2008structural,hirose2014physical}.
This choice is based on the requirements that the material maintain adequate structural properties and low nuclear activation in the presence of high neutron fluxes and high temperatures.
Ferromagnetic materials are also of interest in fusion for reducing magnetic field ripple from toroidal field coils \cite{turner1978, sheffield1993use, kawashima2001demonstration}.
In fusion and in other magnetic systems such as magnetic resonance imaging (MRI), pieces of ferritic material can be used as shims to fine-tune the field \cite{dorri2002passive,lopez2008passive}.
As physics properties of the confined plasma can be highly sensitive to the magnetic field, it is critical to compute the magnetic field including the effects of the magnetic materials, which have motivated studies for both tokamaks \cite{tsuzuki2003high,tobita2003reduction, shinohara2011effects, chiariello2015effective} and stellarators \cite{ikuta1983helical, harmeyer1999effect, ji2017investigation}.
For all these applications, it is desirable to have a fast and flexible method to compute the magnetic field including the ferromagnetic materials, to include in physics calculation workflows.
Moreover, for design optimization, where the field must be evaluated hundreds of times or more, efficiency is a priority, and it is valuable to have derivatives of the field with respect to design variables. 
In the present work, we propose a new calculation method that satisfies these criteria, and demonstrate its use for shape optimization of stellarator electromagnets.

A fundamental challenge in calculating the field in the presence of ferromagnetic materials is that magnetic material in one region affects the magnetization in every other region.
Therefore to achieve consistency, it is necessary to solve a large system of equations.
Furthermore, the relationship between field and magnetization is generally nonlinear, making this system of equations nonlinear, requiring iteration to solve.
Approaches include solving for the field \cite{newman1972gfun}, scalar magnetic potential \cite{carpentier2013resolution}, or vector potential \cite{le2015magnetic} in each element, or using analytic formulas for the field of polyhedra used for the mesh \cite{bjork2021magtense}.
These methods share the advantage that there is no need to mesh the volume outside the magnetic material; only the ferromagnetic domain is meshed.

In contrast to these methods, our approach eliminates the need to solve any system of equations and does not require any iteration.
This simplification is made possible by making two approximations which are well satisfied for fusion systems, as we will show.
The first approximation is that the ferromagnetic materials perturb the total field by a small amount compared to the field from the electromagnets and plasma current.
In several examples shown in the following sections, this approximation is verified to be accurate.
Second, as the saturation magnetization of the material is achieved at field strengths much smaller than the typical fields in magnetic fusion, the material is saturated.
This is true especially for high-field reactor designs based on high-temperature superconductors \cite{creely2020overview, hegna2025infinity, lion2025stellaris}, but is true also for lower-field designs based on low-temperature superconductors.
Together, these approximations mean that the magnetization vector has a known constant magnitude, pointing along the local field computed in the absence of the steel.
Since in this model ferromagnetic material in one region does not affect magnetization in another, no system of equations needs to be solved, and the field can be computed rapidly by straightforward evaluation of Biot-Savart-type integrals without meshing the volume outside the magnetic material.
Specifically our method uses the field from a large number of point dipoles, which conveniently has already been available in several stellarator design frameworks including SIMSOPT \cite{landreman2021simsopt} and FAMUS \cite{zhu2020topology}, so minimal software development was required to implement the method.

Our method differs from previous work on magnetic materials in fusion in several ways.
Our approach is most similar to the one in the code FEMAG \cite{urata2003,shinohara2011effects}, which also makes the same two approximations described above.
However the FEMAG approach also requires that the magnetic material is represented as infinitesmally thin and flat rectangular plates, whereas the approach here allows for arbitrary computer-aided design (CAD) geometry.
Our approach also differs from the previous stellarator analysis in \cite{harmeyer1999effect} which assumed a spatially uniform permeability; that is not an accurate approximation since in fact the magnetization has uniform magnitude due to saturation whereas the field magnitude varies considerably, so their ratio is nonuniform.
Our work can also be contrasted with the related problem of producing stellarator fields with hard permanent magnets \cite{helander2020stellarators, zhu2020topology, hammond2022design, kaptanoglu2023greedy}. 
The permanent magnet problem considered in those references is fundamentally different as the magnetization was considered to be insensitive to the field from electromagnetic coils and plasma current, in contrast to the case of soft ferromagnetic materials considered here.

The computational method presented in this article should be valid for any magnetic fusion concept, including tokamaks, stellarators, and mirror machines, in addition to other high-field devices such as MRI and accelerators.
In the specific case of stellarators, where the shapes and currents of electromagnetic coils are optimized to achieve a desired field in the plasma, the model introduced in this work provides a significant advantage.
As the field contribution from steel can be computed rapidly, it can be included directly in the coil shape and current optimization, which has not been previously demonstrated.
It is straightforward to obtain analytic derivatives of the field from steel with respect to the electromagnet parameters, so the magnet optimization remains gradient-based and efficient.
We show that only small changes to the coil shapes and currents are needed to compensate for the effect of the steel in the blanket.
Therefore, even if significant time is needed for the engineering design of the shapes and locations of the ferromagnetic components, the magnetic field effects can be accounted for by a small adjustment to the coil shapes at the end.

Our mathematical method is described in detail in the following section.
To verify its accuracy, a benchmark with the commercial finite-element code COMSOL \cite{comsol} is presented in section \ref{sec:comsol}, where excellent agreement is demonstrated for the geometry of a modern quasi-isodynamic stellarator reactor.
The coil optimization example is then presented in section \ref{sec:opt}.
In that section, we also demonstrate how the effects of magnetic materials can be included straightforwardly in free-boundary magnetohydrodynamic (MHD) equilibrium calculations, to assess the impact on physics properties such as confinement and stability.
Overall we find that the magnetic materials have little effect on the plasma physics properties of the self-consistent MHD equilibria, with the largest effects being a reduction in rotational transform \cite{harmeyer1999effect,ji2017investigation} that causes a shift in the edge magnetic islands, and a slight destabilization of ideal ballooning modes.
Both effects are compensated by the coil reoptimization that includes the magnetic materials.
Finally in section \ref{sec:conclusion} we discuss the results and conclude.

%%%%%%%%%%%%%%%%%%%%%%%%%%%%%%%%%%%%%%%%%%%%%%%%%%%%%
%%%%%%%%%%%%%%%%%%%%%%%%%%%%%%%%%%%%%%%%%%%%%%%%%%%%%

\section{Efficient calculation of fields from ferromagnetic components}
\label{sec:calculation}

%%%%%%%%%%%%%%%%%%%%%%%%%%%%%%%%%%%%%%%%%%%%%%%%%%%%%

\subsection{Approximations and model equations}
\label{sec:model}

Let $\mathbf{B}^{ferr}$ denote the contribution to the magnetic field due to bound currents in a region of ferromagnetic material.
This quantity at an evaluation point $\mathbf{r}'$ can be computed by an integral of the magnetization density $\mathbf{M}$ over the ferromagnetic region $\Omega$:
\begin{equation}
\mathbf{B}^{ferr}\left(\mathbf{r}'\right)=\frac{\mu_{0}}{4\pi}\int_{\Omega} \frac{d^{3}r}{\left|\mathbf{r}'-\mathbf{r}\right|^{3}}\left[\frac{3\left(\mathbf{r}'-\mathbf{r}\right)\left(\mathbf{r}'-\mathbf{r}\right)\cdot\mathbf{M}\left(\mathbf{r}\right)}{\left|\mathbf{r}'-\mathbf{r}\right|^{2}}-\mathbf{M}\left(\mathbf{r}\right)\right].\label{eq:B_from_M}
\end{equation}
This formula is straightforward to evaluate if $\mathbf{M}$ can be determined.

\begin{figure}[t]
\centering
\includegraphics[width=3in]{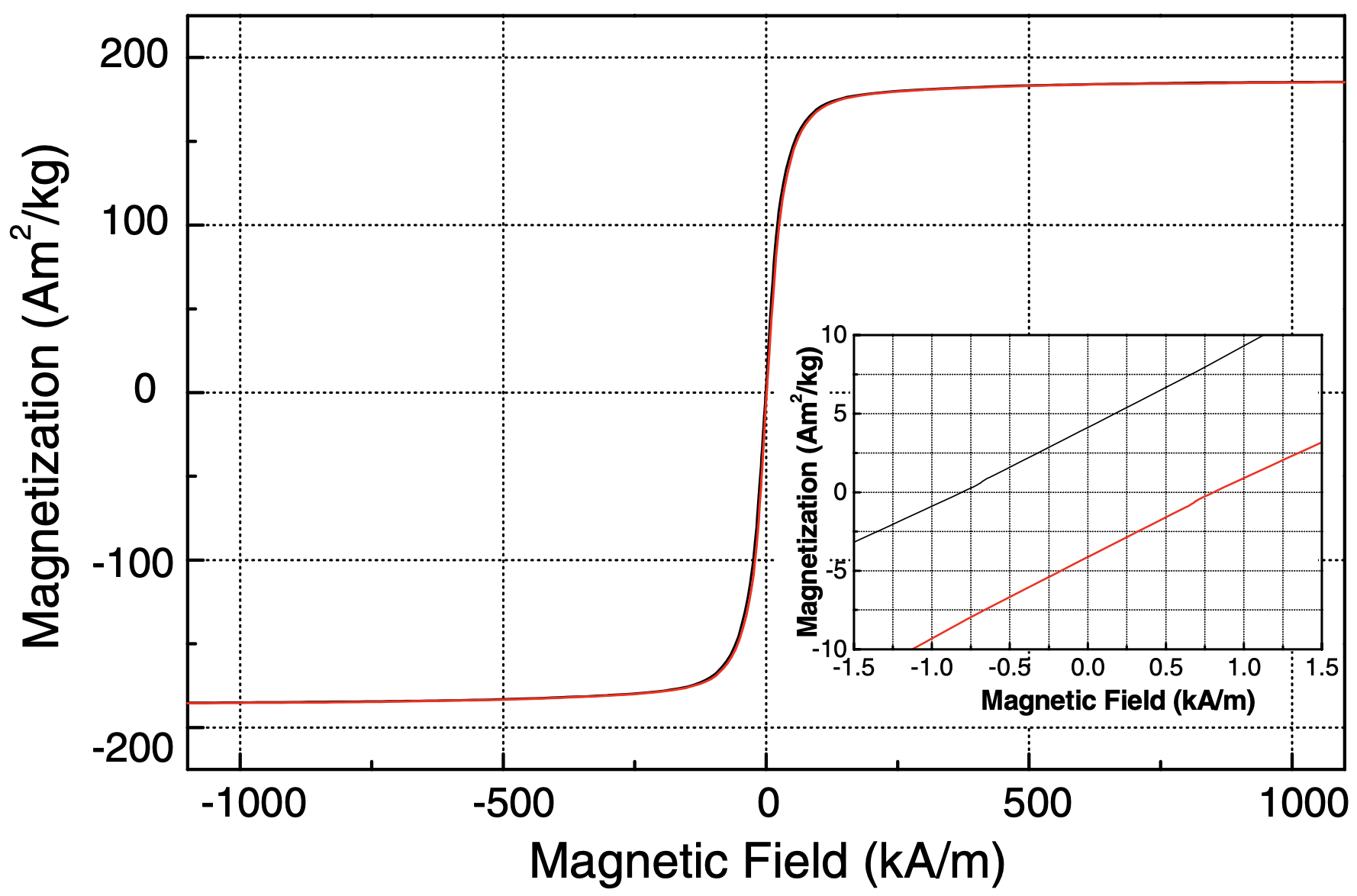}
\caption{
$M-H$ curve for EUROFER97.
Figure reproduced from \cite{mergia2008structural}.
The x-axis range of $\pm 1000$ kA/m is equivalent to $\pm 1.26$ T.
}\label{fig:eurofer_MH}
\end{figure}

To evaluate the magnetization, consider the $M-H$ curve for a typical steel under consideration for fusion blankets, EUROFER97 \cite{mergia2008structural}, shown in figure \ref{fig:eurofer_MH}.
Converting the $H$ values on the x-axis to Tesla by multiplying by $\mu_0$, the range of the x-axis corresponds to $\pm 1.26$ Tesla.
It is apparent that the  magnetization saturates by $\lesssim 0.3 T$.
Typical field strengths in the blanket region of fusion reactor designs are far higher than this, several Tesla or more. 
Thus, the magnitude $M = |\mathbf{M}|$ will be fixed at the saturation value, which we denote $M_{sat}$.
The magnetization is not linearly related to the field so one should not make an approximation of linear permeability.

The saturation magnetization of EUROFER97 is $M_{sat}\sim 1.4\times 10^6$ A/m, so $\mu_0 M_{sat}\sim 1.7$ T.
The value for F82M steel is nearly identical \cite{hirose2014physical}.
This value again is small compared to typical reactor field strengths, particularly for recent high-field designs, so we can approximate $\mu_0 |\mathbf{M}| \ll |\mathbf{B}|$ and $|\mathbf{M}| \ll |\mathbf H|$.

It remains to determine $\mathbf{M}$'s direction.
To do this, we make an ansatz that the the magnetic field contribution from the ferromagnetic material is small compared to the field from the electromagnetic
coils and plasma, $\mathbf{B}^{coils}$ and $\mathbf{B}^{plasma}$ respectively.
Once the total field is evaluated, one can check if this approximation is indeed consistent.
Later in the paper, evidence will be shown that the approximation is in fact highly accurate for typical reactor conditions.
Thus, $\mathbf{M}$ will be aligned with the local magnetic field in the absence of the ferromagnetic material:
\begin{equation}
    \mathbf{M}(\mathbf{r}) \; ||\; (\mathbf{B}^{coils}(\mathbf{r}) + \mathbf{B}^{plasma}(\mathbf{r})).
\end{equation}
The quantity $\mathbf{B}^{plasma}$ can be computed from the Biot-Savart formula by integrating the plasma current over the plasma volume, or by an integral over the plasma boundary via the virtual casing principle \cite{shafranov1972use,drevlak2005pies}.
Note that the evaluation point for $\mathbf{B}^{plasma}$ is strictly outside the plasma, so there is no singular denominator, and sophisticated methods for virtual casing evaluation on the plasma boundary \cite{malhotra2019efficient} are not required.
In sum, the magnitude and direction of $\mathbf{M}$ are known, so eq (\ref{eq:B_from_M}) can be evaluated.

The field from ferromagnetic materials will affect the free-boundary MHD equilibrium, which will affect $\mathbf{B}^{plasma}$ in the ferromagnetic region, which in turn affects $\mathbf{M}$ and hence $\mathbf{B}^{ferr}$.
There are at least four options for addressing this self-consistency issue, which we list here in order of increasing complexity.
First, in many cases, $\mathbf{B}^{plasma}$ in the ferromagnetic region will be significantly smaller than $\mathbf{B}^{coils}$, particularly for quasi-isodynamic stellarators in which the plasma current is very small.
In this case, $\mathbf{B}^{plasma}$ can reasonably be neglected when computing the direction of $\mathbf{M}$.
Second, if $\mathbf{B}^{plasma}$ in the ferromagnetic region is smaller than $\mathbf{B}^{coils}$ but not entirely negligible, one could do the following.
Generate an equilibrium neglecting $\mathbf{B}^{ferr}$, compute $\mathbf{B}^{plasma}$ from this equilibrium, find the resulting $\mathbf{B}^{ferr}$, update the MHD equilibrium to include $\mathbf{B}^{ferr}$, but then not update $\mathbf{B}^{plasma}$ further.
This approach will leave a slight inconsistency between $\mathbf{B}^{ferr}$ and $\mathbf{B}^{plasma}$, but it will be small when $|\mathbf{B}^{plasma}| \ll |\mathbf{B}^{coils}|$.
In a third approach, a fixed-point (Picard) iteration could be done by repeating the steps described in the second method.
This method would be more accurate but also more computationally expensive.
As a fourth option, the evaluation of eq (\ref{eq:B_from_M}) could be included directly in the iterations for the free-boundary MHD equilibrium solve.
Since our motivating application is low-current stellarators, we will focus on the first two iteration-free methods for the remainder of this paper.

%%%%%%%%%%%%%%%%%%%%%%%%%%%%%%%%%%%%%%%%%%%%%%%%%%%%%

\subsection{Quadrature}
\label{sec:quadrature}

We must integrate eq (\ref{eq:B_from_M}) over each volume containing magnetic material.
For numerical calculations, this requires a set of discrete quadrature points $\mathbf{r}_k$ and weights $w_k$, such that volume integrals become $\int_{\Omega} d^3r \, f(\mathbf{r}) \to \sum_k \,  w_k f(\mathbf{r}_k)$.
The quadrature points effectively represent point dipoles.
Here we describe two methods for generating these quadrature grids, using unstructured or structured grids.

The unstructured grid approach is convenient for interfacing with CAD designs, and is required for components with holes such as ports.
Given a CAD description of the steel component, a 3D unstructured mesh is generated.
The barycenter of each mesh element then gives $\mathbf{r}_k$, and the element volume gives the weight $w_k$.
\changed{
Discretization error for this integration rule scales as $\sim O(h^2)$ for element size $h$ \cite{Zienkiewicz2005Finite}.
Higher order quadrature rules over common element types are known and could be used if desired.
}

On other occasions, we have found it useful to use a structured mesh approach to represent toroidally continuous layers.
This allows convenient parameterization of the steel volume, for studies of the effect of plasma-blanket distance or of steel volume, and also provides high-order quadrature accuracy.
The inner toroidal surface is described by a position vector $\mathbf{r}_{in}\left(\theta,\phi\right)$ where $\theta$ and $\phi$ are arbitrary poloidal and toroidal angles.
Similarly, we take the outer bounding surface of each ferromagnetic layer to be described by the position vector $\mathbf{r}_{out}\left(\theta,\phi\right)$.
The surfaces $\mathbf{r}_{in}$ and $\mathbf{r}_{out}$ can be defined by uniform offsets to the plasma surface, or by other means.
A radial coordinate $\rho$ is introduced with range $\left[0,1\right]$ which interpolates between $\mathbf{r}_{in}$ and $\mathbf{r}_{out}$, so that the position vector at any point in the ferromagnetic layer has position vector
\begin{equation}
\mathbf{r}\left(\rho,\theta,\phi\right)=\left(1-\rho\right)\mathbf{r}_{in}\left(\theta,\phi\right)+\rho\mathbf{r}_{out}\left(\theta,\phi\right).\label{eq:position}
\end{equation}
Defining uniform grids in $\theta$ and $\phi$ with spacing $\Delta_\theta$ and $\Delta_\phi$, integration in these angles can be performed using the trapezoid rule with spectral accuracy due to periodicity.
Integration in $\rho$ can be performed using Gauss-Legendre grid points and weights $w_{\rho_k}$.
The overall quadrature weight is then $w_k = \Delta_{\theta} \Delta_{\phi} w_{\rho_k} \sqrt{g}$ where $\sqrt{g} = (\partial \mathbf{r}/\partial\rho)\cdot  (\partial \mathbf{r}/\partial\theta)\times  (\partial \mathbf{r}/\partial\phi)$ is the Jacobian.

In either the unstructured or structured approach, for calculations to be accurate, the distance between quadrature points within the ferromagnetic region should be smaller than the distance to the evaluation point.
The latter is typically the distance to the plasma.
Thus, for components that are closest to the plasma, finer mesh resolution is required.
If the field must be computed very close to the magnetic material, the formula for the field from a uniformly magnetized tetrahedron \cite{nielsen2019stray} can be applied, with the magnetization computed at the element barycenter as in the dipole method.
However we have found that evaluating the field from the simpler dipole formula is sufficient for the applications here.

\begin{figure}[h!]
\centering
\includegraphics[width=\columnwidth]{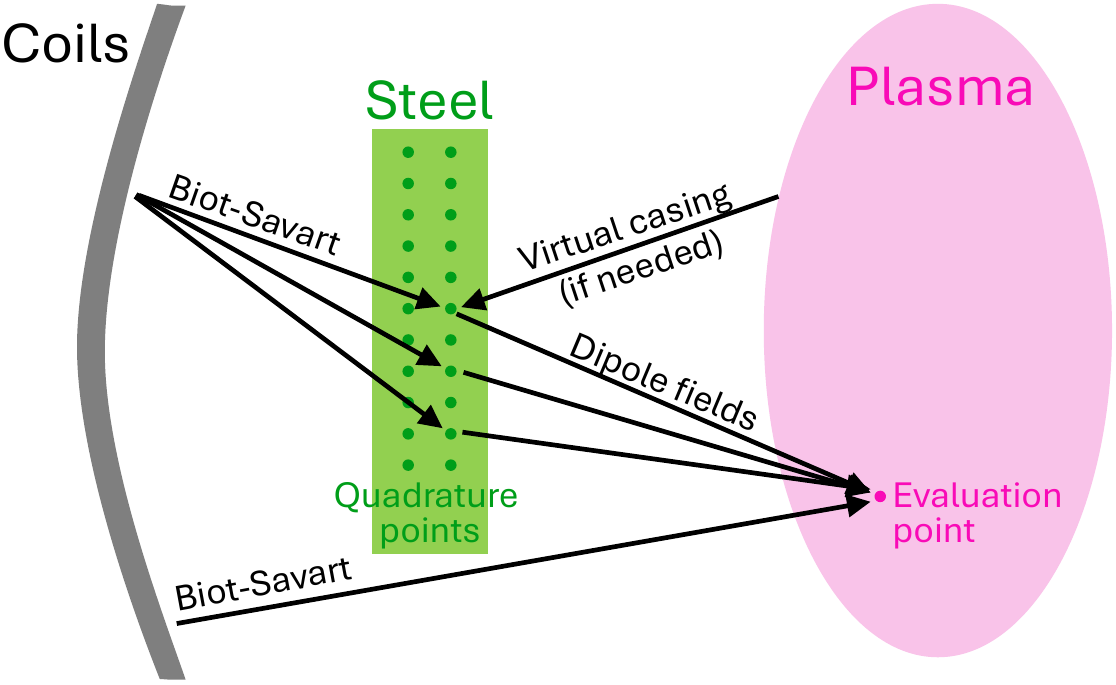}
\caption{
Illustration of the approximate method proposed here, eq (\ref{eq:dipole_sum})-(\ref{eq:dipole_moments}).
The field from the coils is first evaluated at quadrature points in the ferromangetic material.
If the plasma current is substantial, its contribution to the total field there can be added using the virtual casing method.
The field at the quadrature points determines the magnetization there, and the resulting field from the magnetized material can be computed by summing the field of point dipoles at the quadrature points.
This field is computed at the evaluation points of interest, where the direct field from the coils is added.
}\label{fig:illustration}
\end{figure}

To summarize our model, the field contribution from ferromagnetic material is evaluated by a sum of point dipole fields
\begin{equation}
\mathbf{B}^{ferr}\left(\mathbf{r}'\right)=\frac{\mu_{0}}{4\pi} \sum_k  \frac{1}{\left|\mathbf{r}'-\mathbf{r}_k\right|^{3}}\left[\frac{3\left(\mathbf{r}'-\mathbf{r}_k\right)\left(\mathbf{r}'-\mathbf{r}_k\right)\cdot\mathbf{m}_k}{\left|\mathbf{r}'-\mathbf{r}_k\right|^{2}}-\mathbf{m}_k\right],
\label{eq:dipole_sum}
\end{equation}
where the dipole magnitudes are
\begin{equation}
    \mathbf{m}_k = \frac{w_k M_{sat}}{\left| \mathbf{B}^{coils}(\mathbf{r}_k) + \mathbf{B}^{plasma}(\mathbf{r}_k)\right|} \left[\mathbf{B}^{coils}(\mathbf{r}_k) + \mathbf{B}^{plasma}(\mathbf{r}_k)\right],
    \label{eq:dipole_moments}
\end{equation}
and $\mathbf{B}^{plasma}$ may be negligible in the latter formula if the plasma current is sufficiently small.
An illustration of the method is given in figure \ref{fig:illustration}.

%%%%%%%%%%%%%%%%%%%%%%%%%%%%%%%%%%%%%%%%%%%%%%%%%%%%%

\subsection{Forces}
\label{sec:forces}

The method here can also be used to rapidly compute the magnetic force on ferromagnetic components.
The force on a point dipole $\mathbf{m}$ is $\mathbf{f}=(\nabla\mathbf{B}^{ext})\cdot\mathbf{m}$ \cite{griffiths2023introduction}, where $\mathbf{B}^{ext}$ is the field due to sources other than the dipole, so the force density on a magnetized region is $\mathbf{F} =(\nabla\mathbf{B}^{ext})\cdot\mathbf{M}$.
This same formula can also be derived by summing the Lorentz force densities for the volume and boundary of the ferromagnetic region, $\mathbf{F} = \int_{\Omega} d^3r \mathbf{J}_b\times\mathbf{B}^{ext} + \int_{\partial\Omega} d^2r \mathbf{K}_b\times\mathbf{B}^{ext}$, where $\mathbf{J}_b=\nabla\times\mathbf{\mathbf{M}}$ is the current density associated with bound currents and $\mathbf{K}_b=\mathbf{M}\times\mathbf{n}$ is the surface current density associated with bound currents.
This equivalence and other approaches to computing the force are discussed further in the appendix.
The total force on a component follows as
\begin{equation}
    \mathbf{f} = \int_{\Omega} d^3r \,(\nabla\mathbf{B}^{ext})\cdot\mathbf{M},
    \label{eq:force}
\end{equation}
where $\mathbf{B}^{ext}=\mathbf{B}^{coils}+\mathbf{B}^{plasma}$, or simply $\mathbf{B}^{ext} \approx \mathbf{B}^{coils}$ if the plasma current is small.
The volume integral can be discretized using the quadrature methods of the previous section.
Also $\nabla\mathbf{B}^{ext}$ in this expression can be found analytically by differentiating the Biot-Savart law and, if the plasma contribution is included, the virtual casing formula.

%%%%%%%%%%%%%%%%%%%%%%%%%%%%%%%%%%%%%%%%%%%%%%%%%%%%%
%%%%%%%%%%%%%%%%%%%%%%%%%%%%%%%%%%%%%%%%%%%%%%%%%%%%%

\section{Verification with the finite element method}
\label{sec:comsol}

% See lab book
% 20251005-01 B_steel calculations for talk and paper.docx

% Comsol models used:
% 20251007-01_COMSOL_model_of_scaled_config_for_Bsteel_paper_steel.mph
% 20251006-01_COMSOL_model_of_scaled_config_for_Bsteel_paper_vacuum.mph
% 20251007-02_COMSOL_half_period_of_steel_to_compute_force.mph

\begin{figure*}[h]
\centering
\includegraphics[width=2.2in]{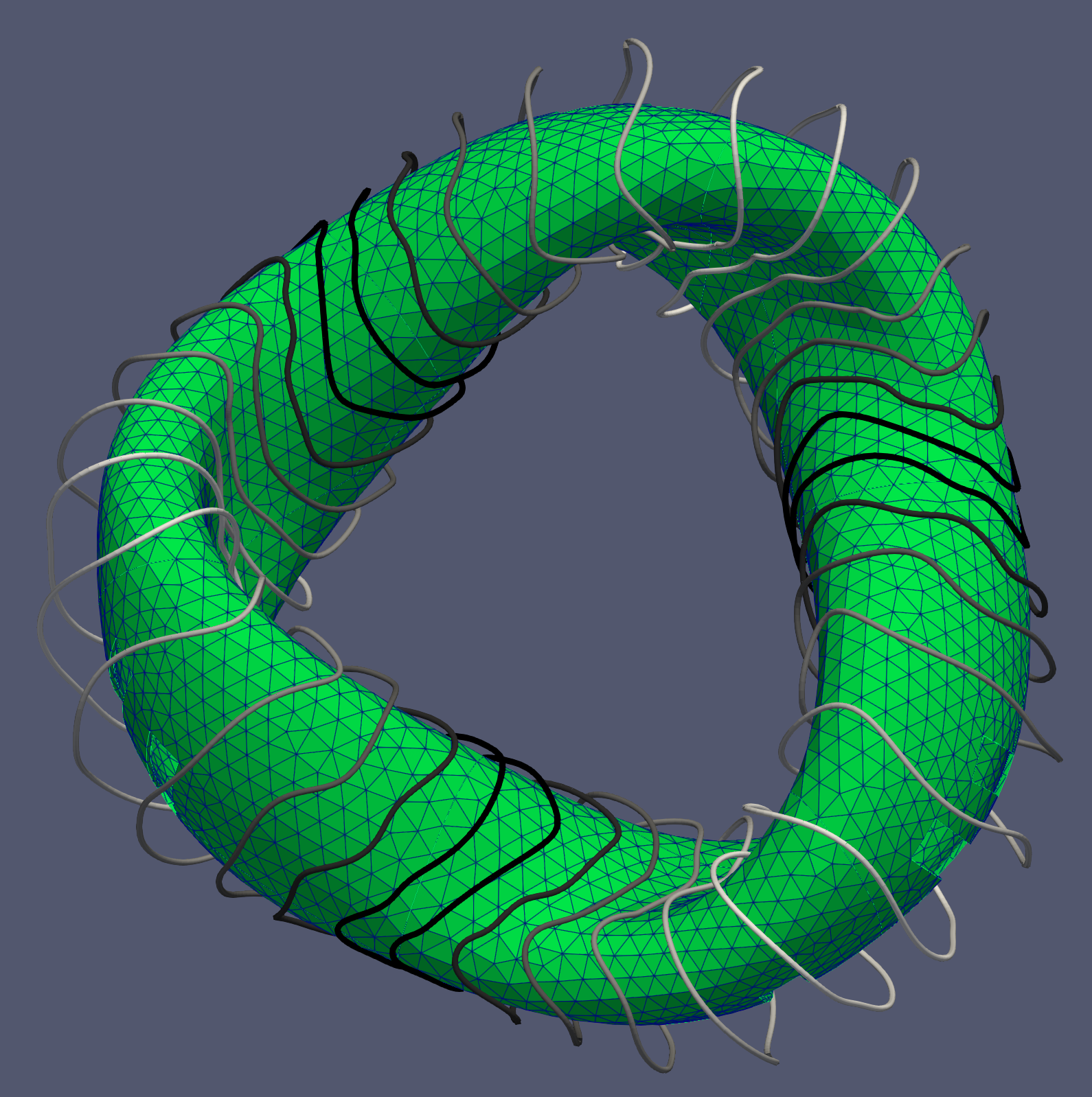}
\includegraphics[width=2.8in]{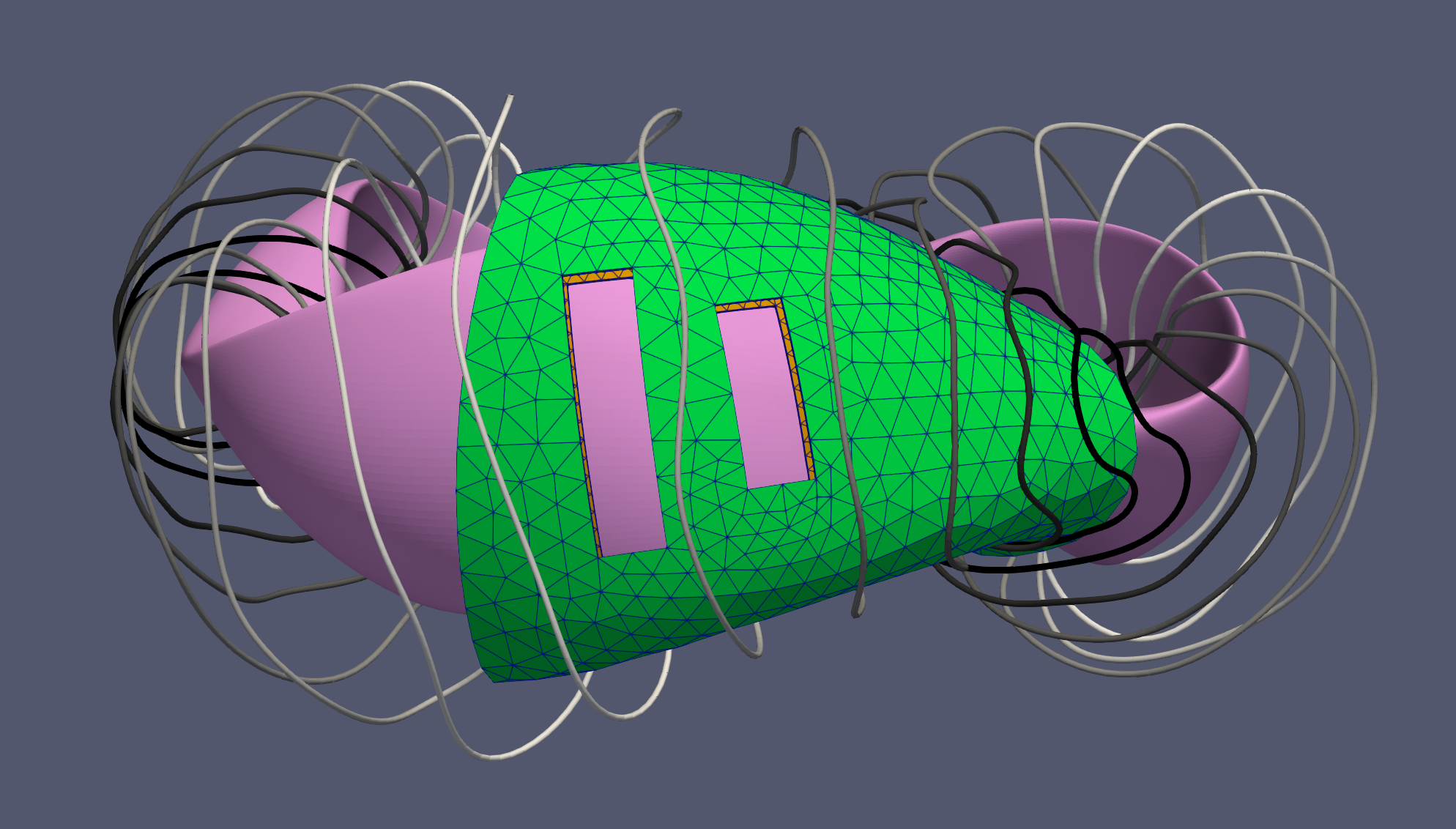}
\includegraphics[width=2.1in]{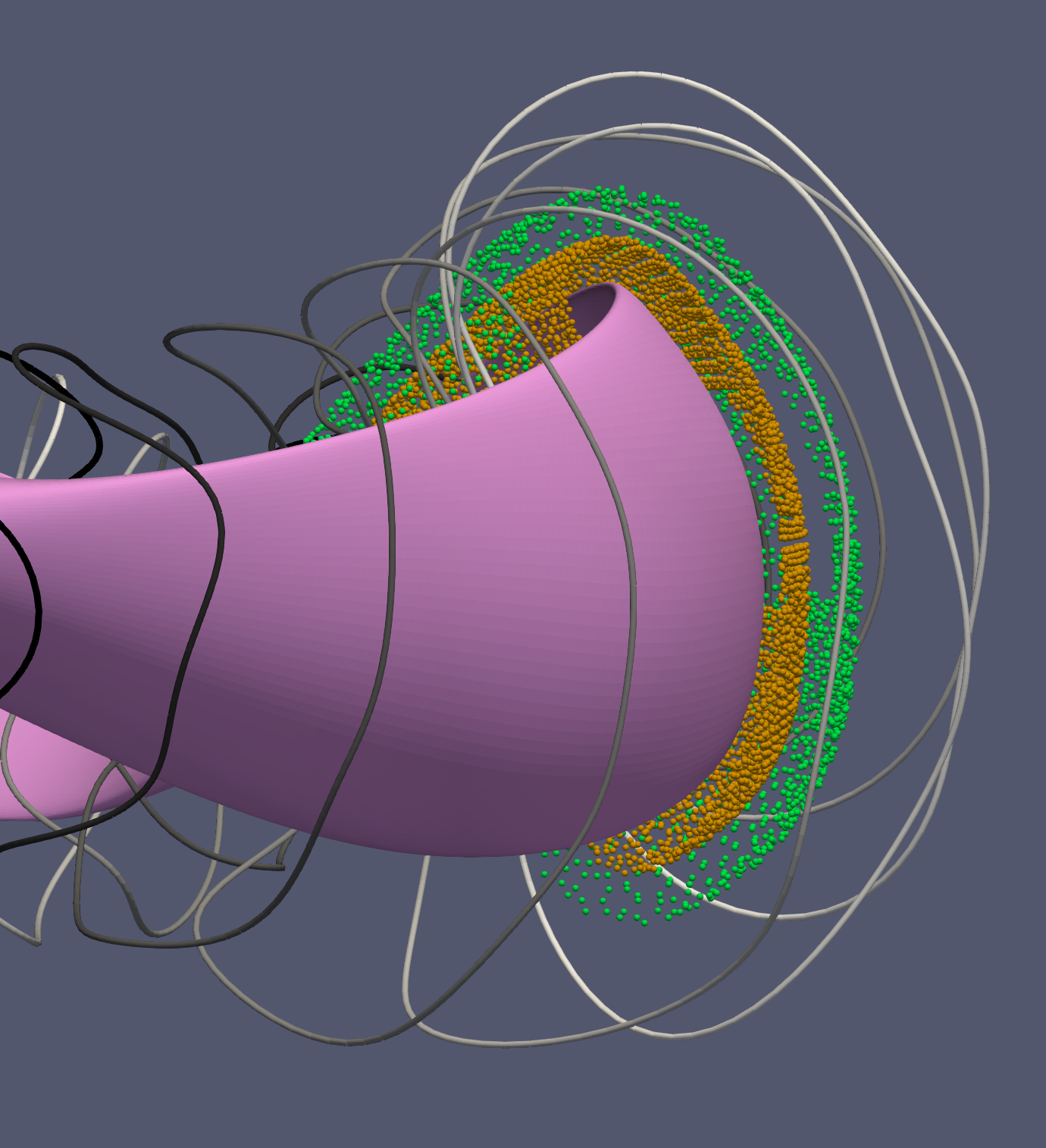}
\caption{
Geometry for the COMSOL benchmark in section \ref{sec:comsol} and optimization example in section \ref{sec:opt_example}.
Left: bird's eye view.
Center: Side view showing the mesh used for the steel layers and the holes for ports.
Right: Side view showing the dipole locations (quadrature points) representing the two steel layers.
In the right two panels, only one half period of the steel region is shown.
}\label{fig:example_setup}
\end{figure*}

To verify that our approximate model is accurate for fusion reactor parameters, we carry out a comparison with the commercial finite-element software COMSOL Multiphysics{\textregistered}\cite{comsol}.
The geometry for the comparison is shown in figure \ref{fig:example_setup}.
The plasma shape, shown in pink, is a three-field-period quasi-isodynamic stellarator.
The configuration is at reactor scale, with average major radius of 12.7 m, and volume-averaged field strength in the plasma of 8 T.
The plasma is confined by a set of 36 modular coils (6 unique shapes each repeated 6 times), shown in gray.
While multi-filament finite-build coils are being used for the design of Type One Energy's stellarators, a single-filament model of the coils is used in this example for simplicity; use of multi-filament coils does not alter any aspect of the method.
Two walls of EUROFER97 are placed between the plasma and coils, one 60 cm from the plasma, and the other 140 cm from the plasma, both 15 cm thick, with holes removed for ports.
These components are not from an actual blanket design, but are merely intended to illustrate a possible application of the method.
Only one half period of the walls is shown in the right two panels, but they extend around the full torus, as in the left panel.
Considering only the field from the coils, the minimum field strength in the plasma region is $>6.4$ T, and the minimum in the steel volume is $>4.3$ T, so the steel is fully saturated.
However the COMSOL calculation uses the full $M(H)$ curve for EUROFER97 without assuming saturation.
COMSOL also fully accounts for the effect of magnetization in one region on the magnetization in another.
\changed{The default second-order elements are used}.
Calculations with the approximate model are performed by first meshing the steel CAD geometry using GMSH \cite{geuzaine2009gmsh}, then applying the open-source SIMSOPT software \cite{landreman2021simsopt}, using its \texttt{DipoleField} class for arrays of point dipoles.
Plasma current is neglected for simplicity in this COMSOL comparison, but will be considered in later sections.

\begin{figure*}[h]
\centering
\includegraphics[width=6.5in]{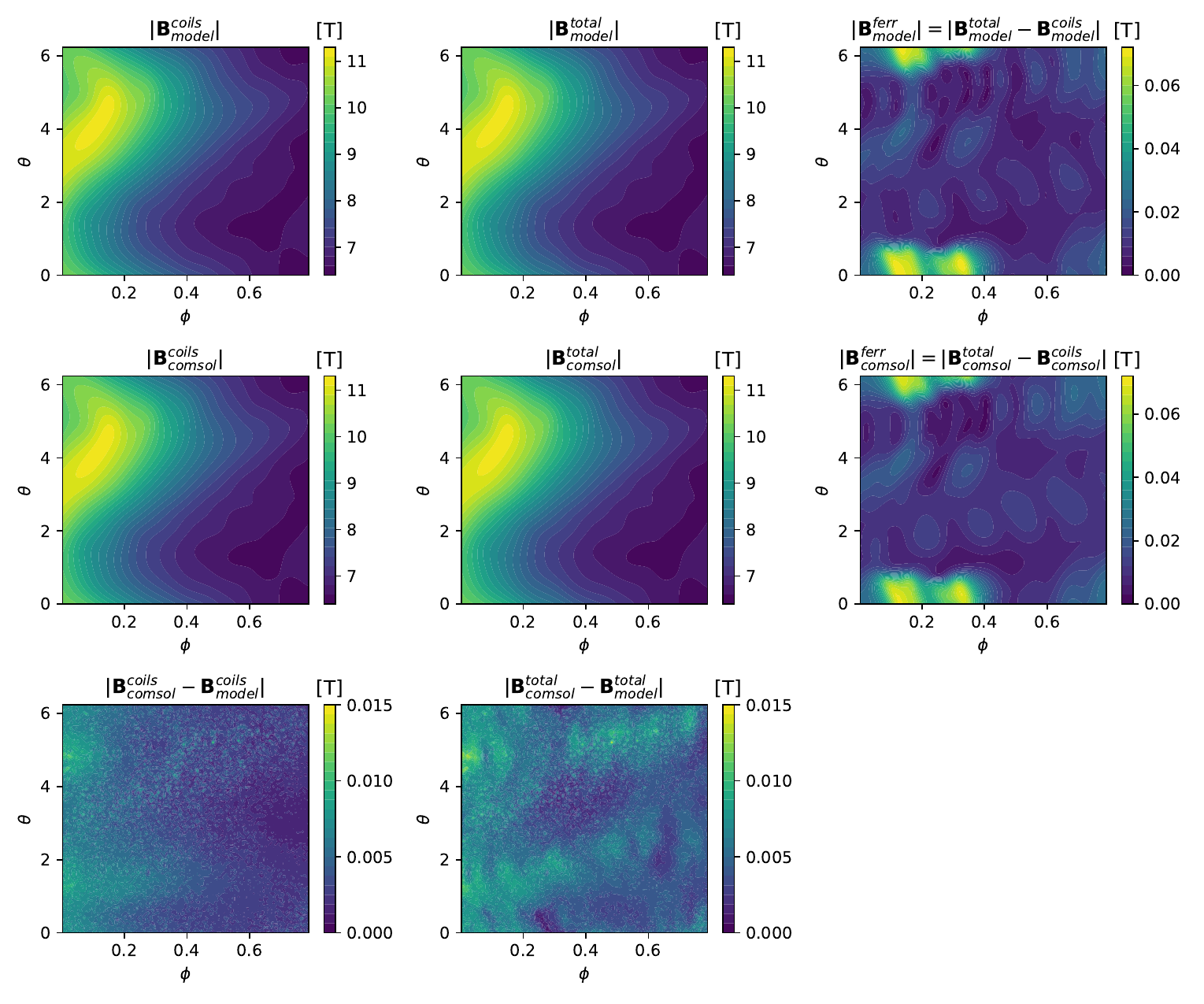}
\caption{
Comparison between our approximate model and a high-fidelity COMSOL calculation for the configuration of figure \ref{fig:example_setup}, showing the magnetic field on the plasma boundary.
Comparing the four panels in the top-left quadrant, both COMSOL and our model agree that the change to $\mathbf{B}$ due to the steel is small compared to the field from the coils.
The change to $\mathbf{B}$ from the steel (right column) agrees between the two methods.
Differences between the two methods (bottom row) are comparable with and without the steel, and are small compared to the effect of the steel (right column).
}\label{fig:comsol}
\end{figure*}

The total field including the steel is computed using both the approximate model and COMSOL.
The field on the plasma boundary is displayed in figure \ref{fig:comsol}.
The four panels in the top-left show the field from only the coils, $\mathbf{B}^{coils}$, and the total field $\mathbf{B}^{total}=\mathbf{B}^{coils} + \mathbf{B}^{ferr}$, for both the model and COMSOL.
Note that the color scale for these four panels is identical.
These four panels are visually indistinguishable, showing both that the effect of the steel is small, and that there is good agreement between the two methods.
The remaining four panels show differences between these fields on \changed{much finer color scales so they can be seen. The right pair of panels share a common color scale, and the bottom pair of panels share a different common color scale.}
The bottom row shows the difference between our model and COMSOL both for the coil field alone and for the total field.
The differences in coil field alone (bottom left) are due only to discretization error since the underlying equations are identical in the absence of steel.
The differences in total field (bottom center) arise due to both discretization error and the different underlying equations.
However the discrepancy in the bottom center panel is not much larger than the bottom left panel, indicating that the physics differences are comparable to discretization error.
The right two panels show $|\mathbf{B}^{ferr}|$ computed by the two approaches, showing close similarity.
The maximum $|\mathbf{B}^{ferr}|$ occurs under the vertical edges of the ports.

% Timing in ~/work25/20251005-01_B_steel_analysis_for_scaled_mattland_20250106_289/20251005-01-010_BSteel_for_coils_from_002/timing
To evaluate the magnetic field for figure \ref{fig:comsol}, COMSOL took 409 seconds on a 2023 M2 Macbook Pro.
For comparison, generating the data for figure \ref{fig:comsol} with the reduced model took only 0.33 seconds on the same computer, giving a factor of $>1000\times$ speed-up.

\begin{figure}[h]
\centering
\includegraphics[width=\columnwidth]{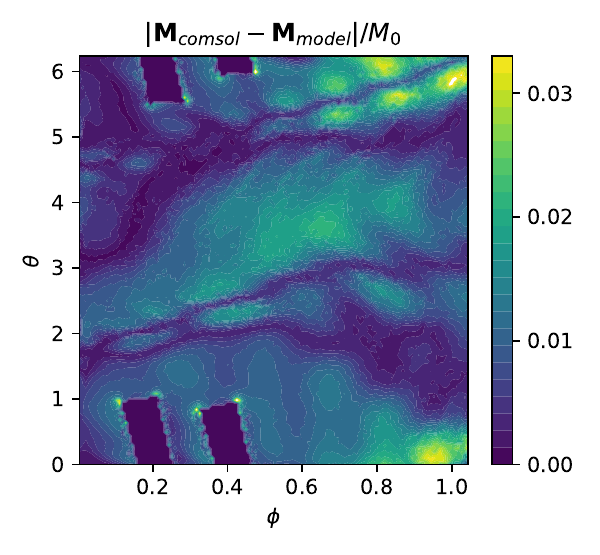}
\caption{
The difference in the magnetization vector between COMSOL and the approximate model, normalized to the magnitude of magnetization in the approximate model, is small.
Data are evaluated in the middle of the inner steel layer.
}\label{fig:magnetization_vs_comsol}
\end{figure}

We can also directly assess the accuracy of the approximation used to determine the magnetization in the reduced model.
This is done in figure \ref{fig:magnetization_vs_comsol}, which shows the norm of the difference in magnetization vectors between COMSOL and the reduced model.
The magnetization is evaluated on a surface 0.675 m from the plasma, in the middle of the inner steel layer.
\changed{In the ports, visible as the rectangular regions on the top and bottom left of the figure, the difference is necessarily zero.}
The difference is normalized to the magnitude of the magnetization in the reduced model, i.e. the saturation magnetization.
It can be seen that this relative difference in magnetization is no more than a few percent, meaning the approximation is quite good.

\begin{figure}[h]
\centering
\includegraphics[width=\columnwidth]{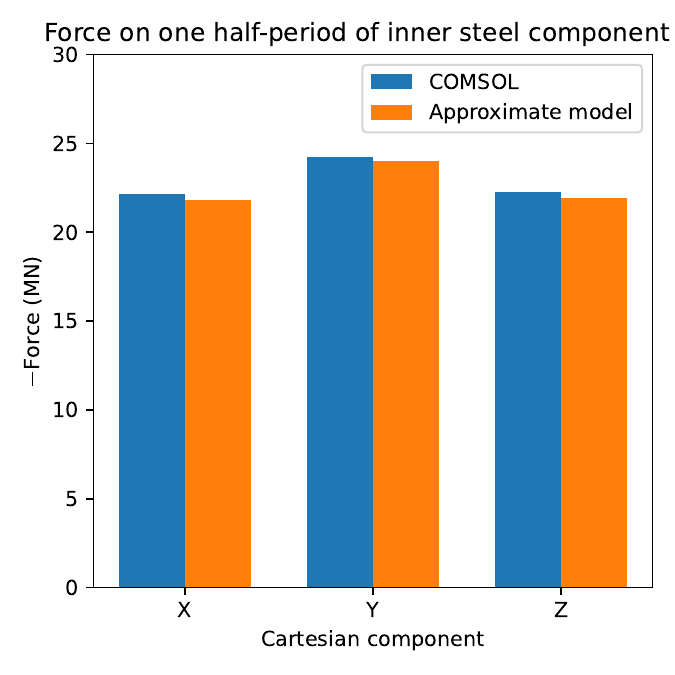}
\caption{
The approximate model presented in this article also provides accurate calculations of the forces on steel components. 
}\label{fig:force}
\end{figure}

\begin{figure}[h]
\centering
\includegraphics[width=\columnwidth]{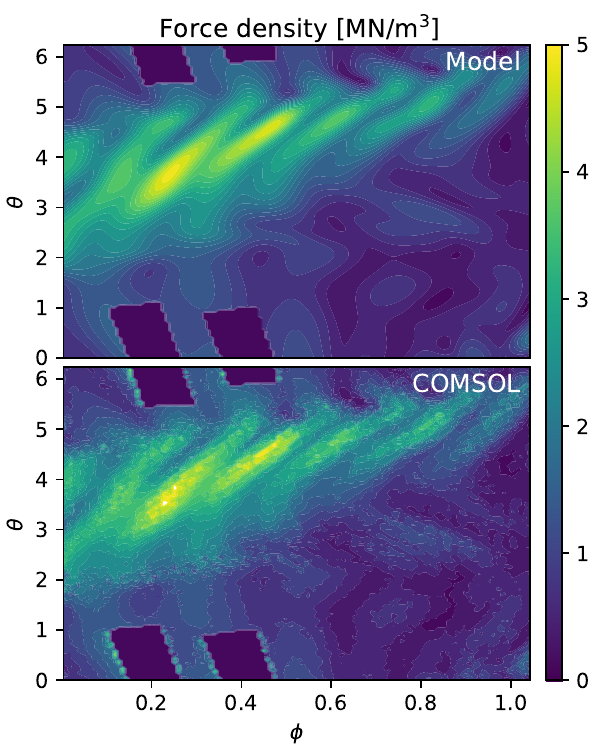}
\caption{
\changed{
The approximate model presented in this article also provides the force density within steel components, showing general agreement with COMSOL.
}
}\label{fig:force_density}
\end{figure}

The forces on the steel can also be compared between the two methods.
In COMSOL the force is calculated from a surface integral of the Maxwell stress on the boundary instead of the volume integral (\ref{eq:force}).
For this comparison we evaluate the net force on one half field period of the inner steel region.
Results are displayed in figure \ref{fig:force}.
The Cartesian components of this force in MN were found to be [-22.1, -24.2, -22.3] from COMSOL, and [-21.8, -24.0, -21.9] according to the reduced method introduced in this article, based on eq \ref{eq:force}).
Thus, good agreement is obtained.
\changed{
Agreement is also found for the volumetric force density within the steel, the magnitude of which is shown in figure \ref{fig:force_density}.
As with figure \ref{fig:magnetization_vs_comsol}, data are shown in the middle of the inner steel layer.
Of the various approaches that can be considered to compute magnetic forces (\ref{app1}), 
the Kelvin formula $\mu_0 \mathbf{M}\cdot\nabla\mathbf{H}$ \cite{Melcher, bobbio2000equivalent} is used to compute the force density in COMSOL.
This quantity is somewhat noisy since the finite element solution for the magnetic field must be numerically differentiated.
However COMSOL and the approximate model generally show agreement in the magnitude and spatial pattern of the force density.
}

%%%%%%%%%%%%%%%%%%%%%%%%%%%%%%%%%%%%%%%%%%%%%%%%%%%%%
%%%%%%%%%%%%%%%%%%%%%%%%%%%%%%%%%%%%%%%%%%%%%%%%%%%%%

\section{Stellarator coil optimization}
\label{sec:opt}

%%%%%%%%%%%%%%%%%%%%%%%%%%%%%%%%%%%%%%%%%%%%%%%%%%%%%

\subsection{Overview}
\label{sec:opt_overview}

Next we demonstrate that the approximate model can be applied straightforwardly to include ferromagnetic components in stellarator coil optimization.
In this example we consider the second stage of the standard two-stage approach to stellarator design.
A target plasma equilibrium (from stage 1) is given, and the task is to optimize the shapes and currents of electromagnets to support that equilibrium.
This is done by minimizing the component of the actual magnetic field normal to the target plasma boundary surface, while satisfying engineering constraints on the coils such as a maximum allowed curvature and minimum clearances.

Since the effects of ferromagnetic materials are minor, as shown in figure \ref{fig:comsol} and as will be shown in more detail below, coils can initially be designed without considering magnetic materials.
Next, the ferromagnetic components associated with the blanket, first wall, and divertor can be designed.
This step may be a labor-intensive process involving detailed engineering design.
Then, a few iterations of optimization are carried out that include the ferromagnetic components.
The coil parameters will only slightly change in this step.
The change may be sufficiently small that no design modification of the ferromagnetic components is required.
If some adjustments are required, they should be small.
A few more iterations of coil optimization could then be carried out, and the process should converge rapidly.

%%%%%%%%%%%%%%%%%%%%%%%%%%%%%%%%%%%%%%%%%%%%%%%%%%%%%

\subsection{Derivatives}
\label{sec:derivative}

The efficiency of optimization is significantly improved when derivative information is available, which is usually the case for stellarator coil optimization.
Let us therefore examine the relevant derivatives involving the ferromagnetic material to ensure they can be computed.
Usually the objective function for coil design is based on the quantity $f = \int_{\partial P} d^2a (\mathbf{B}^{total} \cdot\mathbf{n})^2$, where the integral is over the desired surface shape of the plasma $\partial P$ and $\mathbf{n}$ is the unit normal.
The gradient of this quantity with respect to the coil shapes and current is required.
Let $x$ denote one of these design variables, so the derivatives required for optimization are $\partial f / \partial x$.
Since the objective $f$ depends on $\mathbf{B}^{total}$ and hence on $\mathbf{B}^{ferr}$, then we need $\partial \mathbf{B}^{fer} / \partial x$, where the numerator is evaluated at any of the fixed set of points on the target plasma surface.
If the coils do not evolve much when the ferromagnetic material is introduced, then $\mathbf{B}^{fer}$ will be approximately constant, so $\partial \mathbf{B}^{fer} / \partial x$ could be neglected.
However it is straightforward to compute $\partial \mathbf{B}^{fer} / \partial x$, so it may as well be retained.
To compute this derivative, we can write
\begin{equation}
\frac{\partial \mathbf{B}^{ferr}}{\partial x}
 = \sum_{i,j,k}
 \frac{\partial \mathbf{B}^{ferr}}{\partial m_i(\mathbf{r}_k)}
 \frac{\partial m_i(\mathbf{r}_k)}{\partial B^{coils}_j(\mathbf{r}_k)}
  \frac{\partial B^{coils}_j(\mathbf{r}_k)}{\partial x}.
  \label{eq:chain_rule}
\end{equation}
Here, $i$ and $j$ are sums over the three Cartesian components, $m_i$ and $B^{coils}_j$ are the components of the dipole moments $\mathbf{m}$ and coil field $\mathbf{B}^{coils}$, and $k$ is a sum over the dipole locations.
The effect of $\mathbf{B}^{plasma}$ on the dipole moments does not appear in (\ref{eq:chain_rule}) because $\mathbf{B}^{plasma}$ is considered fixed during the optimization, as is normally done when evaluating the gradient of $f$.
The first term, $\partial \mathbf{B}^{ferr} / \partial m_i(\mathbf{r}_k)$, is readily evaluated from (\ref{eq:dipole_sum}):
\begin{equation}
\frac{\partial B^{ferr}_{\ell} \left(\mathbf{r}'\right)}{\partial m_i(\mathbf{r}_k)}
=\frac{\mu_0}{4\pi \left|\mathbf{r}'-\mathbf{r}_k\right|^{3}}\left[\frac{3\left(\mathbf{r}'-\mathbf{r}_k\right)_i \left(\mathbf{r}'-\mathbf{r}_k\right)_\ell}{\left|\mathbf{r}'-\mathbf{r}_k\right|^{2}}- \delta_{i,\ell} \right],
\end{equation}
for each component $\ell$. The middle term of (\ref{eq:chain_rule}) is evaluated from (\ref{eq:dipole_moments}):
\begin{eqnarray}
 \frac{\partial m_i}{\partial B^{coils}_j} 
 &=& \frac{w_k M_{sat}}{\left| \mathbf{B}^{coils} + \mathbf{B}^{plasma}\right|} \\
&&\times \left[\delta_{i,j} - \frac{(B^{coils}_i + B^{plasma}_i)(B^{coils}_j + B^{plasma}_j)}{\left| \mathbf{B}^{coils} + \mathbf{B}^{plasma}\right|^2} \right], \nonumber
\end{eqnarray}
where all quantities are evaluated at $\mathbf{r}_k$.
The last term of (\ref{eq:chain_rule}), $\partial B^{coils}_j(\mathbf{r}_k) / \partial x$, is already routinely computed for coil optimization (albeit with the numerator evaluated on the plasma surface rather than at the dipole locations), so we assume this term can be calculated.
Thus all the factors in (\ref{eq:chain_rule}) can be evaluated.
To get the gradient of the scalar objective $f$, it is most efficient to proceed from left to right (in ``reverse mode'') using vector-Jacobian products.

For the optimization example in the next section, we find that neglecting $\partial\mathbf{B}^{ferr}/\partial x$ in the optimization does not reduce the convergence rate, since $|\mathbf{B}^{ferr}|$ is so much smaller than $|\mathbf{B}^{coils}|$, and dropping this term reduces the time per iteration.
However in cases with lower field strength where $\mathbf{B}^{ferr}$ is relatively more significant, retaining this term may be important.

%%%%%%%%%%%%%%%%%%%%%%%%%%%%%%%%%%%%%%%%%%%%%%%%%%%%%

\subsection{Example}
\label{sec:opt_example}

\begin{figure*}[h]
\centering
\includegraphics[width=7.2in]{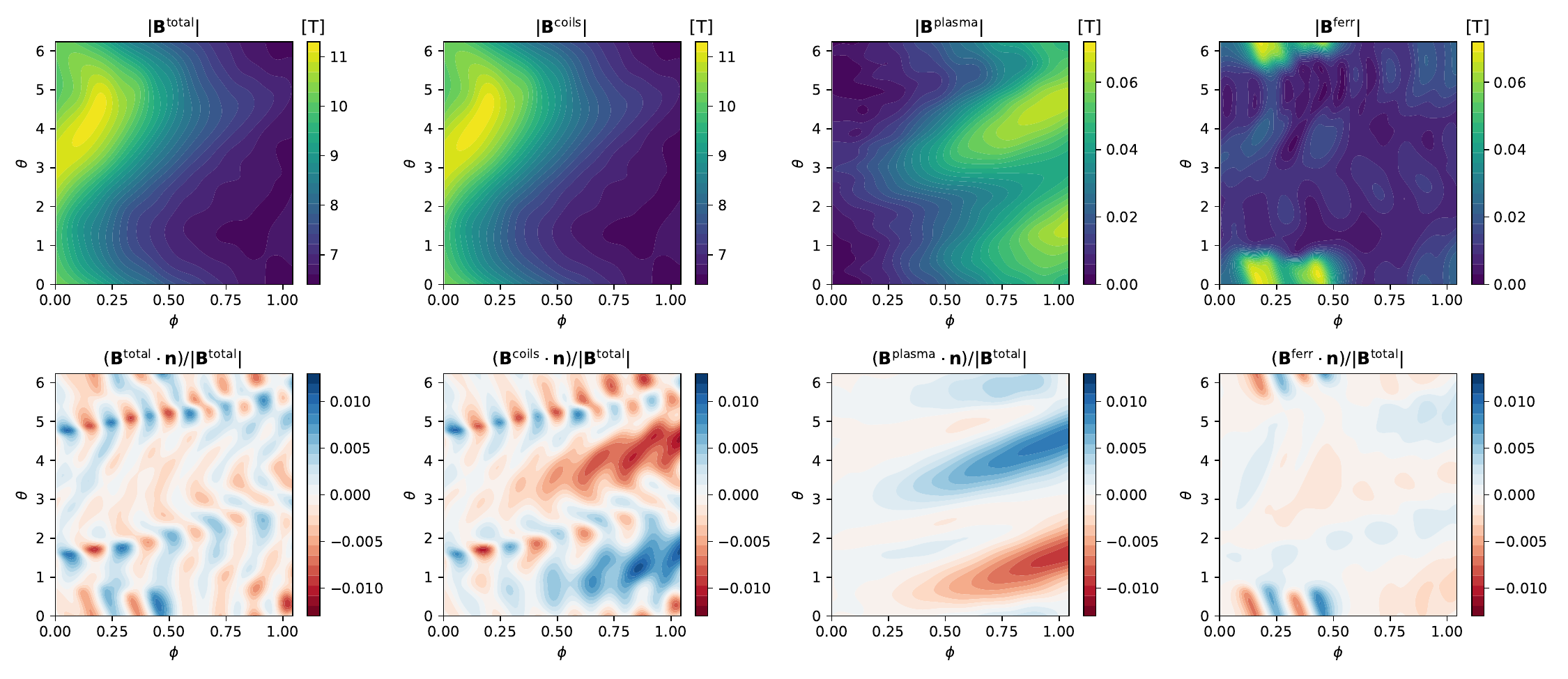}
% Generated using field_from_steel.plots.plot_for_paper()
\caption{
Contributions to the magnetic field and its component normal to the target plasma surface.
Data here are for the initial coils shown in figure \ref{fig:example_setup}.
}\label{fig:initial_B}
\end{figure*}

To illustrate the method with an example, we return to the quasi-isodynamic stellarator and steel geometry shown in figure \ref{fig:example_setup}, and explained in section \ref{sec:comsol}.
The magnetic field for the initial state is displayed in figure \ref{fig:initial_B}.
The eight panels all show aspects of the magnetic field on the target plasma surface, i.e. the surface that was used in the initial optimization of the coil shapes in which the ferromagnetic components were neglected.
The plasma configuration has a ratio of plasma pressure to magnetic pressure of $\beta=1.7\%$, so the field due to the plasma current is non-negligible compared to coil ripple on the plasma surface, and is computed using the virtual casing method \cite{malhotra2019efficient}.
However, throughout the steel region, the ratio $|\mathbf{B}^{plasma}| / |\mathbf{B}^{coils}|$ is found to be quite small, $<2.4\%$, so $|\mathbf{B}^{plasma}|$ is neglected for determining the steel magnetization in the rest of this paper.
Indeed, the effect of plasma current on the steel field would be a small correction to an already small correction.
% This value 2.4% was computed using 20251005-01-compute_Bplasma_over_Bcoils.py
The four columns show the total field and the separate contributions from coils, plasma current, and the steel.
The top row shows the magnitude of the field in Tesla, while the bottom row shows the field component normal to the target plasma surface, as this normal component highlights the errors in producing the desired flux surface.
It can be seen that $\mathbf{B}^{ferr}$ is quite small, less than 1\% of the total field.
This fact supports our approximation that $\mathbf{B}^{ferr}$ can be neglected when computing the direction of the magnetization, and suggests that the effect of $\mathbf{B}^{ferr}$ on the plasma properties is unlikely to be dramatic.
Nonetheless, looking at the components normal to the target plasma surface in the bottom row, $\mathbf{B}^{ferr}\cdot\mathbf{n}$ is comparable to the modular coil ripple and $\mathbf{B}^{plasma}\cdot\mathbf{n}$.
Therefore $\mathbf{B}^{ferr}$ should not be ignored completely, and we can expect it may be large enough to have some effect on certain physics properties of the plasma.
As noted above, the largest magnetic perturbation from the steel is in the region \changed{corresponding to} the ports.

Free-boundary MHD equilibria are computed for these initial coils, both without and with the steel.
The equilibria are computed using the code VMEC \cite{hirshman1986three}.
For the free-boundary calculation with $\mathbf{B}^{ferr}$, the field from the dipoles representing the steel is evaluated on a tensor-product grid in cylindrical coordinates and saved in VMEC's standard MGRID input format.

\begin{figure}[h]
\centering
\includegraphics[width=\columnwidth]{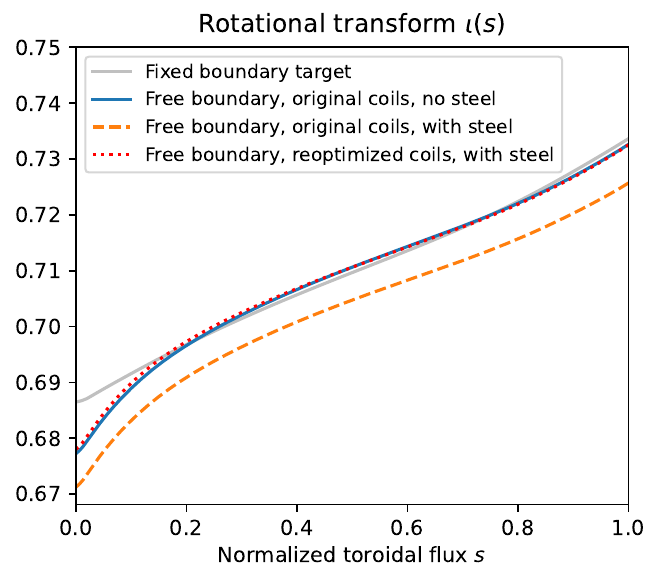}
\caption{
Rotational transform profiles for different equilibria associated with the plasma configuration in figure \ref{fig:example_setup}.
}\label{fig:iota}
\end{figure}

Figure \ref{fig:iota} shows the profiles of rotational transform $\iota$ for the original fixed-boundary target configuration as well as for free-boundary equilibria both excluding and including the field from the steel.
If the steel field is excluded, the coils do a good job matching the $\iota$ profile of the target configuration.
However when the steel field is included, there is a systematic decrease in the $\iota$  profile.
This $\iota$ downshift due to ferromagnetic material has been noted previously \cite{harmeyer1999effect,ji2017investigation}.
It can be understood from analysis of a straight cylinder, where the ferromagnetic layers reduce the poloidal field while not affecting the toroidal field \cite{harmeyer1999effect}.

\begin{figure}[h!]
\centering
\includegraphics[width=\columnwidth]{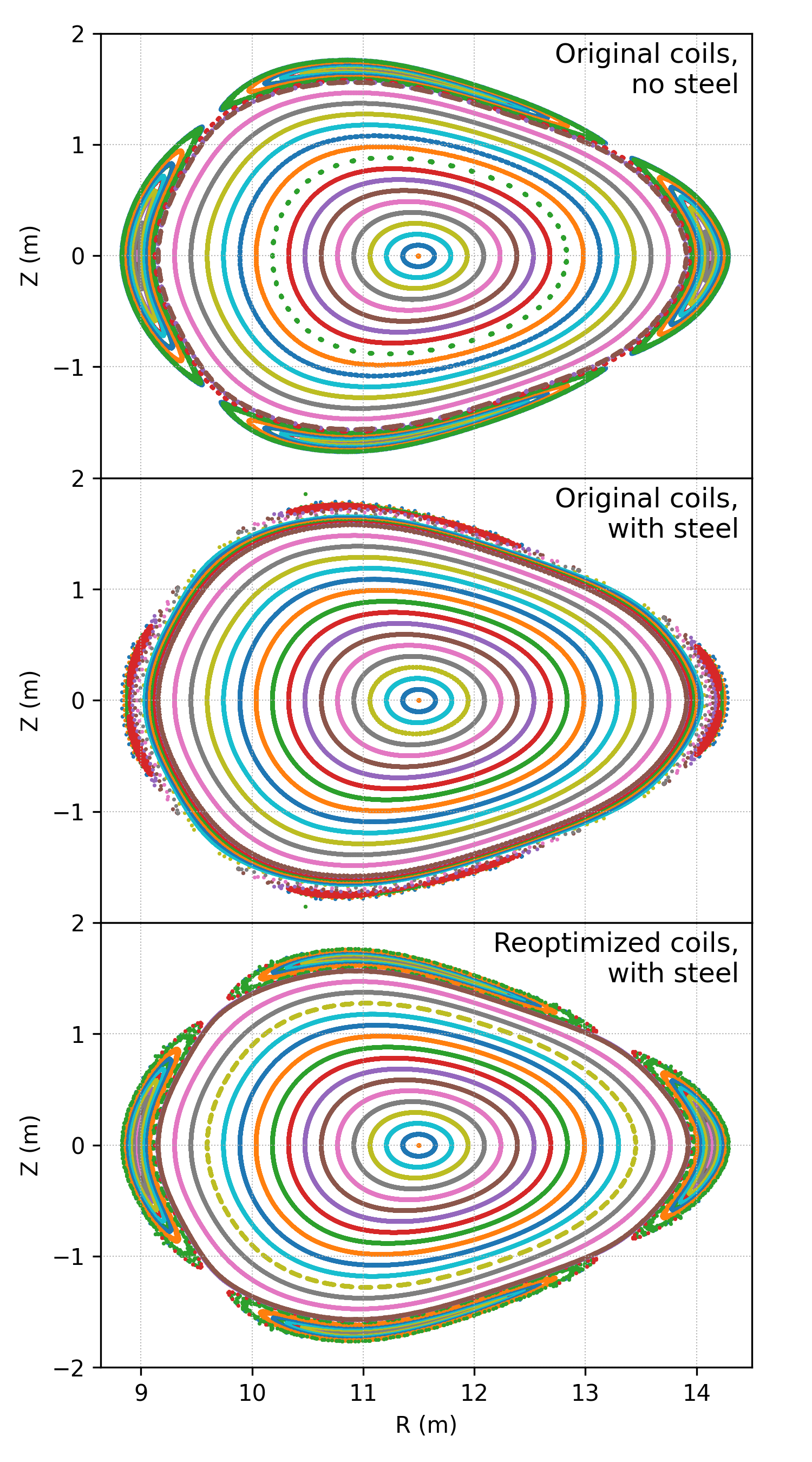}
\caption{
The downshift in $\iota$ due to the steel causes an outward shift of the edge island chain, in addition to distortion from harmonic resonances.
The island location and shape is restored by minor reoptimization of the coils including the fields from the steel.
The same initial locations for field line tracing are used in all three panels.
}\label{fig:islands}
\end{figure}

For stellarator designs with an island divertor, such as W7-X and the configuration here (which uses the $\iota=3/4$ resonance just outside the plasma), this shift to $\iota$ will affect the locations of heat strike points on the divertor, so it must be accounted for.
This effect is seen in figure \ref{fig:islands}, which show Poincare plots at the operating $\beta = 1.7\%$ computed using the code HINT \cite{suzuki2006development, suzuki2017hint}.
The Poincare plots at $\beta=0$ ($\mathbf{B}^{coils} + \mathbf{B}^{ferr}$ only) are nearly identical.
The original coils were designed to produce a chain of four islands at the plasma edge (top panel).
When the field from the steel is included (middle panel), the islands are shifted outward from the domain.
Also a small higher-harmonic island has arisen due to the field perturbation.
While both of these issues could be corrected with additional control coils, we can address them directly with small adjustments to the modular coil shapes.

\begin{figure}[h]
\centering
\includegraphics[width=\columnwidth]{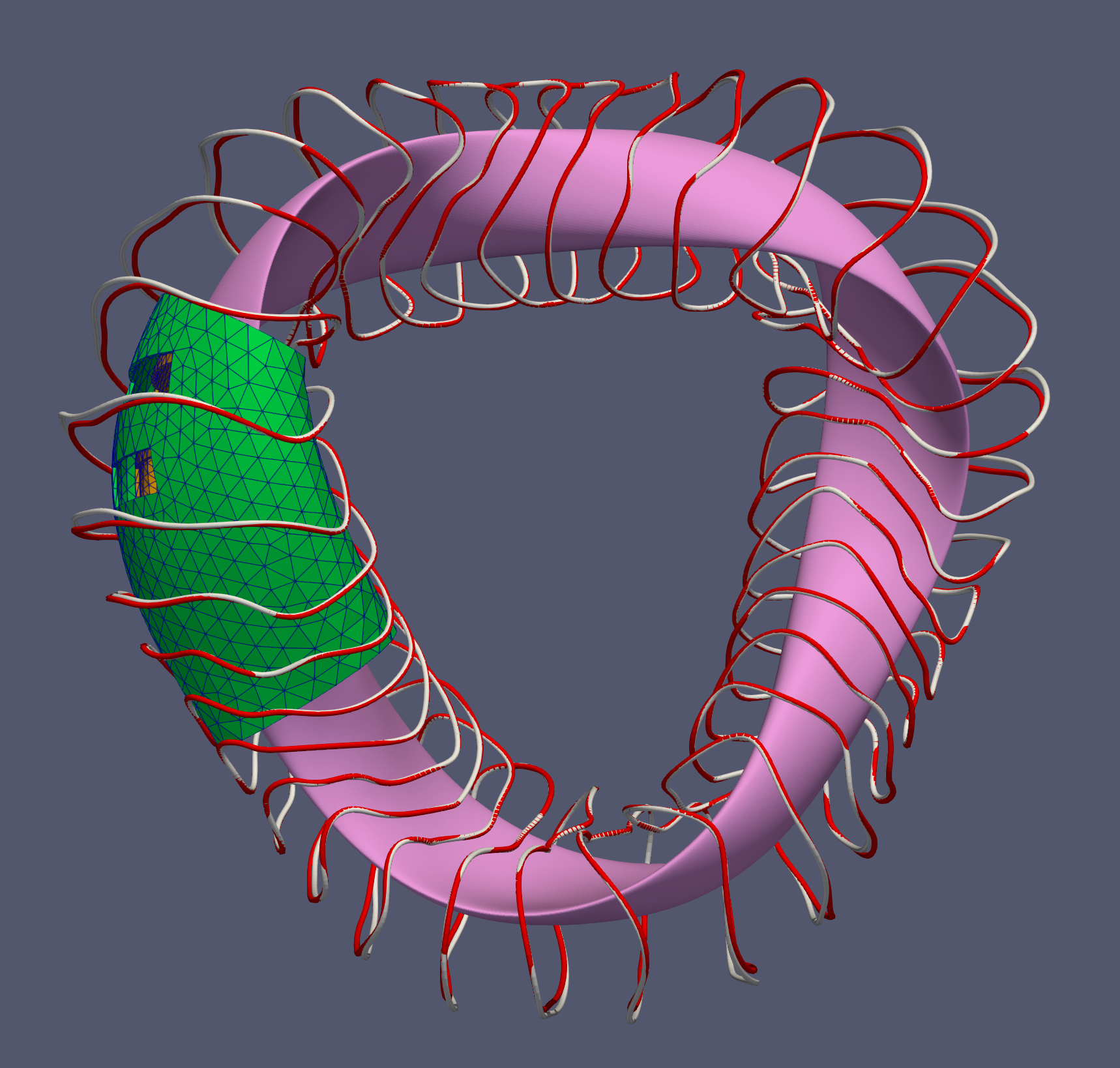}
\caption{
Coils before (white) and after (red) the optimization that includes $\mathbf{B}^{ferr}$.
}\label{fig:compare_coils}
\end{figure}

To compensate for the field from the steel, derivative-based coil optimization is then run to adjust the coil shapes.
The objective function minimized is conventional, with terms to produce the desired plasma shape and control the coil length, coil-coil distance, coil-plasma distance, maximum curvature, and islands respectively:
\begin{eqnarray}
\label{eq:objective}
f &=& \frac{1}{2}\int_{\partial P} d^2a \frac{\left(\mathbf{B}^{total}\cdot\mathbf{n}\right)^2}{B^2} + \left(L_* - \sum_{j=1}^N L_j\right)^2 \\
&&+ \sum_{j=1}^N \sum_{k=1}^{j-1} \int_j d\ell \int_k d\ell \; \max\left(0, \; d_{cc*} - |\mathbf{r}_j - \mathbf{r}_k|\right)^2
\nonumber \\
&&+ \sum_{j=1}^N \int_j d\ell \int_{\partial P} d^2a \;\max\left(0, \; d_{cs*} - |\mathbf{r}_j - \mathbf{r}_S|\right)^2 \nonumber \\
&&+ \frac{1}{2} \sum_{j=1}^N \int_j
d\ell \; \max\left( 0, \; \kappa - \kappa_*\right)^2 \nonumber \\
&&+ \sum_{k=1}^3 w_k \left(b^{res}_k - b^{res}_{k*}\right)^2. \nonumber
\end{eqnarray}
Here, $\int_{\partial P} d^2a$ is a surface integral over the plasma boundary, $N$ is the number of unique coil shapes, $j$ is an index over these shapes, $L_j$ is the length of coil $j$, $\int_j d\ell$ is an integral over coil $j$ with respect to arclength, $\kappa(\ell)$ is the curvature, $\mathbf{r}_j(\ell)$ is the position vector along coil $j$, and $\mathbf{r}_S$ is a position vector on this surface.
The quantity $b^{res}_k$ is the $k$-th harmonic resonant field, eq (14) in \cite{boozer2004physics}, associated with the edge island chain:
\begin{equation}
\changed{
b_k^{res}=\frac{1}{2\pi^2} \int d\vartheta \int d\phi \frac{\mathbf{B}\cdot\nabla\psi}{\mathbf{B}\cdot\nabla\phi}\sin(k[4\vartheta-3\phi]),}
\end{equation}
\changed{for toroidal flux $\psi$ and straight-field-line poloidal angle $\vartheta$.
In (\ref{eq:objective}), the second and third harmonics ($k=2,3$) are controlled in addition to the fundamental ($k=1$) to prevent the appearance of higher-order island chains.}
The quantities $L_*$, $\kappa_*$, $d_{cc*}$, $d_{cs*}$, and $b^{res}_{k*}$ are constant target values, and are set to the same values used to generate the original coils in figure \ref{fig:example_setup}.
Therefore the coil lengths, maximum curvature, minimum coil-coil distance, and minimum coil-plasma distance are practically unchanged during the optimization that includes the steel.
Other than the weights $w_k$ on the resonances, no additional weights in (\ref{eq:objective}) are required.
The coils before and after the optimization are compared in figure \ref{fig:compare_coils}.
It can be seen that the changes to the coil shapes are relatively minor, which is not surprising given that the field from the steel is small compared to the total field.
The most noticeable changes to the coil shapes are in the regions on the outboard midplane near the ports.

\begin{figure*}[h]
\centering
\includegraphics[width=7.2in]{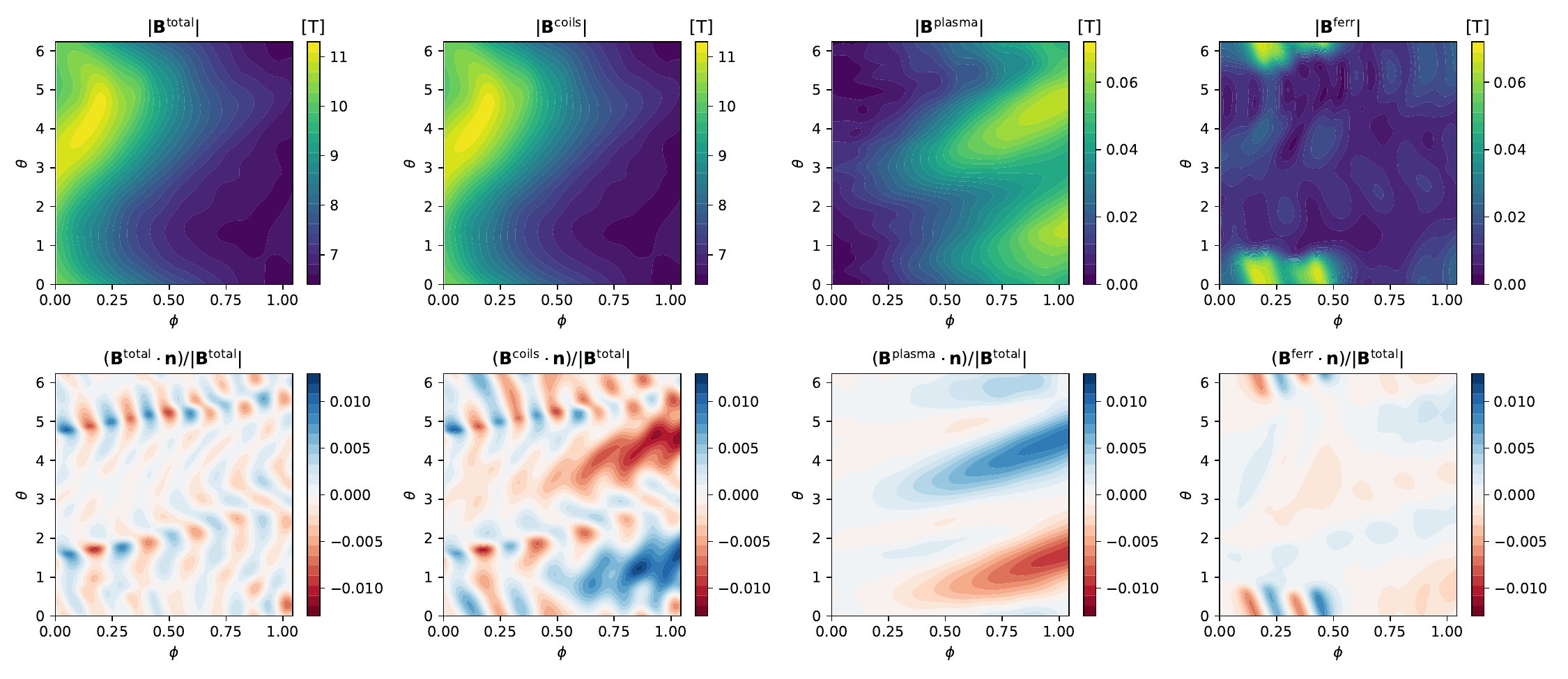}
% Generated using field_from_steel.plots.plot_for_paper()
\caption{
Contributions to the magnetic field and its component normal to the target plasma surface.
Data here are for the coils after optimization to include $\mathbf{B}^{ferr}$.
}\label{fig:final_B}
\end{figure*}

Figure \ref{fig:final_B} shows the contributions to $\mathbf{B}$, and their components normal to the target plasma boundary, following the optimization.
Comparing figures \ref{fig:initial_B} and \ref{fig:final_B}, the right columns show that $\mathbf{B}^{ferr}$ is nearly unchanged, reflecting the fact that the change to the coil shapes is small.
Looking at the sixth panel ($\mathbf{B}^{coils}\cdot\mathbf{n})/|\mathbf{B}^{total}|$, focusing on the lower left corner, it can be seen that the optimized coils generate an approximately equal and opposite ripple to the ripple from the steel near the ports.
As a result, comparing the lower-left panels of figures \ref{fig:initial_B} and \ref{fig:final_B}, the total ripple near the ports is significantly reduced.

A free-boundary equilibrium calculation is run using the optimized coils and including $\mathbf{B}^{ferr}$.
The resulting profile of rotational transform is included in figure \ref{fig:iota}.
It can be seen that the coil optimization including $\mathbf{B}^{ferr}$ eliminates the downshift in $\iota$, making it nearly indistinguishable from the free-boundary solution with the original coils without the steel.
The bottom panel of figure \ref{fig:islands} shows that the modified coils also restore the edge island chain to the desired location and shape.

\begin{figure}[h]
\centering
\includegraphics[width=\columnwidth]{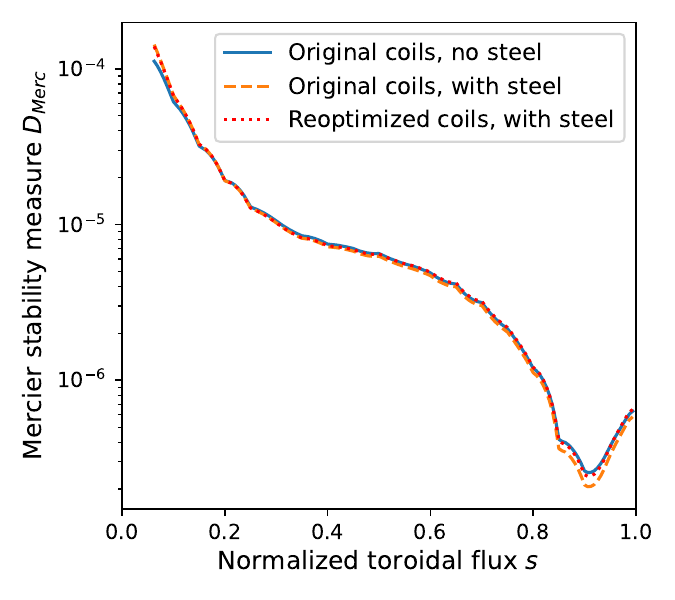}
\caption{
The steel is found to have negligible impact on the Mercier criterion for MHD stability.
Positive values indicate stability.
}\label{fig:Mercier}
\end{figure}

\begin{figure}[h]
\centering
\includegraphics[width=\columnwidth]{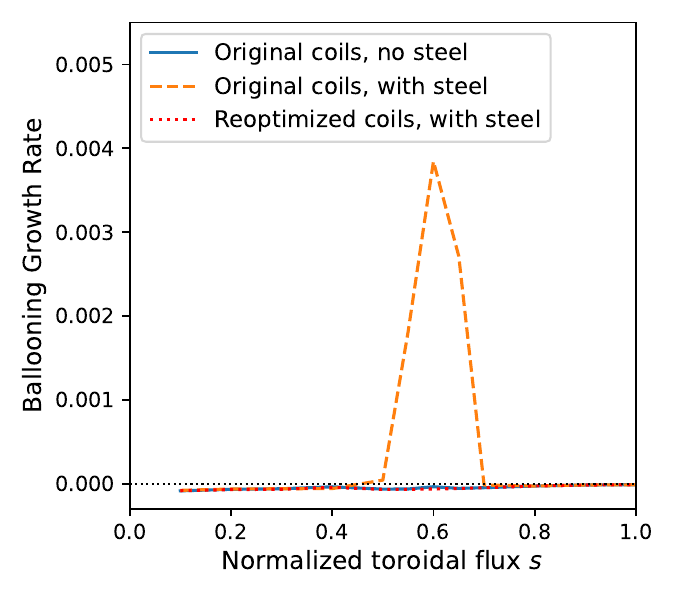}
\caption{
The steel is found to slightly destabilize ideal MHD ballooning modes, but this issue is corrected when the coils are reoptimized to acount for the steel.
}\label{fig:ballooning}
\end{figure}

From the free-boundary equilibria that include the steel field, we can also examine the effect of the steel on other stability and confinement properties.
Figure \ref{fig:Mercier} shows the Mercier stability measure $D_{Merc}$ for the original coils without and with the steel, and for the reoptimized coils.
Positive value indicate stability.
The ferritic material has no significant effect.
The effect on ideal ballooning stability is shown in figure \ref{fig:ballooning}.
At each radius, stability calculations are carried out for field lines centered on many poloidal and toroidal locations, and the maximum growth rate is shown, so positive values indicate instability.
When the steel field is added to the original coils, ballooning modes are slightly destabilized, although the growth rate is small.
The reoptimized coils restore ballooning stability, by better matching the target plasma configuration.

\begin{figure}[h]
\centering
\includegraphics[width=\columnwidth]{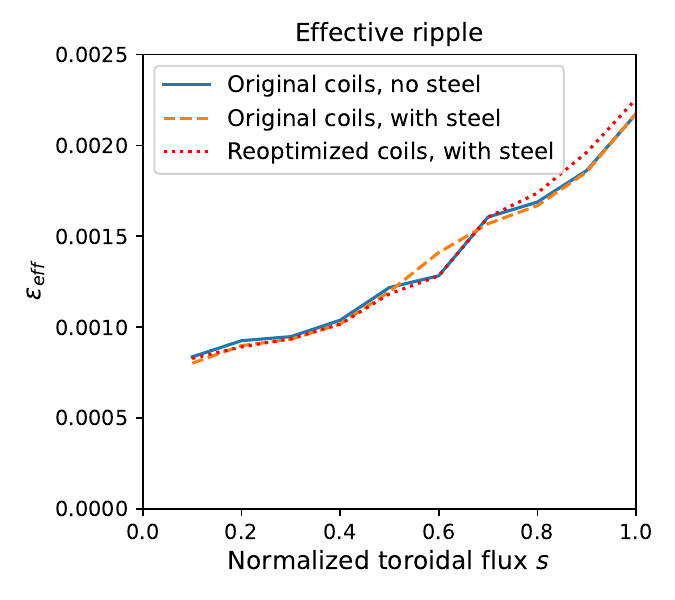}
\caption{
The steel is found to have no significant effect on the neoclassical transport metric $\epsilon_{eff}$ (effective ripple).
}\label{fig:eps_eff}
\end{figure}

\begin{figure}[h]
\centering
\includegraphics[width=\columnwidth]{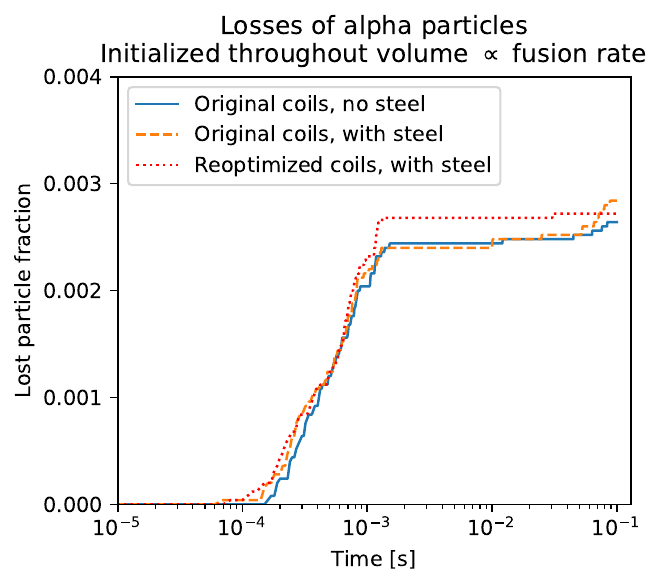}
\caption{
\changed{
The steel is found to not have a significant effect on confinement of guiding-center trajectories.
}
}\label{fig:simple_volume}

\end{figure}

\begin{figure}[h]
\centering
\includegraphics[width=\columnwidth]{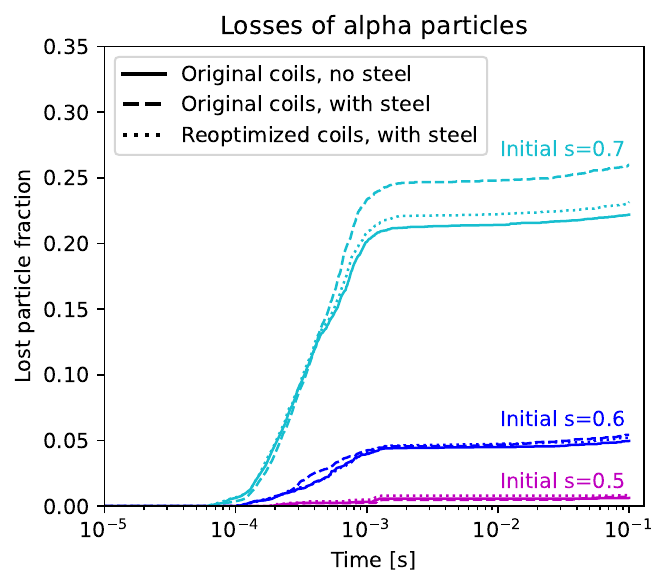}
\caption{
\changed{
Even for guiding center trajectories initialized close enough to the plasma edge to be lost, the steel is found to have only a minor effect on confinement.}
}\label{fig:simple_single_surface}
\end{figure}

We also find that the field from the ferritic material has negligible effect on several measures of confinement.
The neoclassical transport measure $\epsilon_{eff}$ is shown in figure \ref{fig:eps_eff} for the three free-boundary configurations.
No systematic difference is apparent.
Next, figure \ref{fig:simple_volume} shows the results of guiding-center trajectory calculations in the three free-boundary configurations.
For these calculations, conducted using the code SIMPLE \cite{albert2020accelerated, albert2020symplectic}, \changed{first 25,000} particles are initialized 
\changed{throughout the plasma volume proportional to the fusion reaction rate from the local density and temperature,}
and followed without collisions for 0.1 seconds.
The original size and field strength of the configurations are used for the tracing.
\changed{
In all three configurations, losses are extremely low, $< 0.3\%$, without any significant effect from the steel.
It is conceivable that differences might be greater if particles are initialized at larger minor radius, closer to the steel.
To check this hypothesis, further calculations are done with particles initialized on specific flux surfaces, with results shown in figure \ref{fig:simple_single_surface}.
}
No particles are lost if the initial surface $s$ (normalized toroidal flux) is $<0.5$, and losses from $s=0.5$ are $<0.9\%$.
For particles initialized farther out ($s$= 0.6-0.7), losses may be slightly higher for the original coils with steel, but the effect is small.
Overall, we conclude that the field from the steel components does not significantly affect \changed{either the effective ripple $\epsilon_{eff}$ or energetic particle confinement}.


%%%%%%%%%%%%%%%%%%%%%%%%%%%%%%%%%%%%%%%%%%%%%%%%%%%%%
%%%%%%%%%%%%%%%%%%%%%%%%%%%%%%%%%%%%%%%%%%%%%%%%%%%%%

\section{Discussion and conclusion}
\label{sec:conclusion}

In this work we have demonstrated a method to efficiently account for the magnetic fields from ferromagnetic materials in systems with strong electromagnets, such as fusion reactors.
In this case, when the field from the ferromagnetic materials is small compared to that from the coils, the magnetization is nearly aligned with the coil field.
This approximation eliminates the need to solve a large system of equations to account for the effect of magnetization in one region on another, significantly accelerating calculations.
Instead, all that is required is evaluating the Biot-Savart formula for the coil field inside the material, and then evaluating the formula for the field of dipoles representing the material.
Moreover in fusion systems the magnetic material is fully saturated, so the magnetization's magnitude is a constant, further simplifying calculations.
The method also enables rapid calculation of the forces on ferromagnetic components.
By comparisons to COMSOL calculations that fully account for interactions of the material with itself, we have shown that the approximations in our model are extremely well satisfied for modern stellarator reactor parameters.
We expect the method will be useful and accurate for tokamaks as well, and for other $\gtrapprox 1$ Tesla devices such as magnetic resonance imaging systems.

As the method requires only evaluating the field of an array of dipoles, with no complicated solvers, it is straightforward to include in fusion device design workflows.
Here we have demonstrated including the field from ferritic blanket components in free-boundary MHD equilibrium calculations.
Thus it was possible to assess the effect of the steel components on plasma confinement and stability.
Even for a significant volume of steel with large edges that create magnetic ripple, the effects on the physics properties were found to be quite minor.
The primary effects were the previously-known decrease in rotational transform, which can be important for an island divertor, and a slight destabilization of ideal ballooning modes.

We also demonstrated that the magnetic effects of the steel can be included in stellarator optimization.
Differentiating through the model is straightforward, preserving the efficiency of gradient-based optimization.
In the example here, the electromagnet shapes were optimized, but in principle the location and shape of steel components could be optimized as well in the same framework.
Since only a small adjustment to the coil shapes was needed to compensate for the steel, it may not be necessary to adjust the steel location when the coils are adjusted.
This is advantageous since design of the blanket components may be a labor-intensive process.
Through optimization of the electromagnets to account for the steel, we demonstrated it was possible to restore the edge island chain and ballooning stability.
We expect that the methods here will thus be valuable and effective for optimizing designs of the next generation of fusion facilities.

%%%%%%%%%%%%%%%%%%%%%%%%%%%%%%%%%%%%%%%%%%%%%%%%%%%%%
%%%%%%%%%%%%%%%%%%%%%%%%%%%%%%%%%%%%%%%%%%%%%%%%%%%%%

\section*{Declaration of competing interest}

The work of M.L. was performed as a consultant and was not part of the employee’s responsibilities to the University of Maryland.

%%%%%%%%%%%%%%%%%%%%%%%%%%%%%%%%%%%%%%%%%%%%%%%%%%%%%
%%%%%%%%%%%%%%%%%%%%%%%%%%%%%%%%%%%%%%%%%%%%%%%%%%%%%

\section*{Acknowledgments}

This research was funded by Type One Energy.
We gratefully acknowledge assistance with CAD geometry from Bonita Goh, Dan Theilen, Kevin Rogers, and Ian Harnett, guidance from Tate McCartney regarding the HINT code, and conversations about this work with Chris Hegna, John Canik, Daniel Clark, Madelynn Knilans, Michael Zarnstorff, and Dominic Seidita.
This work was also enabled by the computational tools developed and maintained by the Type One Energy optimization group.

%%%%%%%%%%%%%%%%%%%%%%%%%%%%%%%%%%%%%%%%%%%%%%%%%%%%%
%%%%%%%%%%%%%%%%%%%%%%%%%%%%%%%%%%%%%%%%%%%%%%%%%%%%%

\appendix
\section{Force on a magnetized region}
\label{app1}

Forces on magnetized objects can be computed in a wide variety of ways: integration of the Maxwell stress over a surface, differentiating an energy (giving the so-called Korteweg–Helmholtz force), equivalent magnetic charges, or the Lorentz force on magnetization currents or dipoles \cite{Melcher,deMedeiros1999distribution,engel2002electromagnetic,jackson,bobbio2000equivalent}.
Some of these methods can be expressed either in terms of the total magnetic field or the part from sources outside the magnetized domain \cite{bobbio2000equivalent}.
Here we demonstrate the equivalence of two intuitive methods for computing the force, the magnetization current and dipole methods.
%present an alternative derivation of the force on a magnetized region, eq (\ref{eq:force}).
Let $\mathbf{f}$ denote the total force on a magnetized region $\Omega$.
This force can be understood to originate from the Lorentz force on the magnetization currents:
\begin{equation}
    \mathbf{f} = \int_{\Omega} d^3r \, \mathbf{J}_b\times\mathbf{B} + \int_{\partial\Omega} d^2r \, \mathbf{K}_b\times\mathbf{B},
    \label{eq:force_lorentz}
\end{equation}
where $\mathbf{J}_b=\nabla\times\mathbf{\mathbf{M}}$ is the current density associated with bound currents and $\mathbf{K}_b=\mathbf{M}\times\mathbf{n}$ is the surface current density associated with bound currents, with $\mathbf{n}$ the outward unit normal of $\Omega$.
In (\ref{eq:force_lorentz}), $\mathbf{B}$ can be taken to be $\mathbf{B}^{ext}$, the field from only sources outside $\Omega$, namely $\mathbf{B}^{coils}+\mathbf{B}^{plasma}$, since the field from the magnetized material cannot exert any net force on itself \cite{jackson,landau}.
To simplify notation the ext superscript is omitted in this appendix. 
We apply $(\nabla\times\mathbf{M})\times\mathbf{B}=-\nabla(\mathbf{M}\cdot\mathbf{B})+(\nabla\mathbf{B})\cdot\mathbf{M}+\mathbf{B}\cdot\nabla\mathbf{M}$ to the first term of (\ref{eq:force_lorentz}), and apply the BAC-CAB rule to the second term.
Using $\int_{\Omega} d^3r \nabla g = \int_{\partial\Omega}d^2r\,\mathbf{n}g$ for $g=\mathbf{M}\cdot\mathbf{B}$, a cancellation occurs to leave
\begin{equation}
\mathbf{f}=   \int_{\Omega} d^3r\left[(\nabla\mathbf{B})\cdot\mathbf{M}+\mathbf{B}\cdot\nabla\mathbf{M}\right] -\int_{\partial\Omega} d^2r\, \mathbf{n}\cdot\mathbf{B}\mathbf{M}.
\label{eq:appendix_almost_done}
\end{equation}
Next we apply $\int_{\Omega} d^3r \nabla\cdot \overleftrightarrow{\mathbf{T}} = \int_{\partial\Omega} d^2r \,\mathbf{n}\cdot \overleftrightarrow{\mathbf{T}}$ (true for any tensor $\overleftrightarrow{\mathbf{T}}$) for $\overleftrightarrow{\mathbf{T}}=\mathbf{B}\mathbf{M}$.
This identity, with $\nabla\cdot\mathbf{B}=0$, implies $\int_{\partial\Omega} d^2r\, \mathbf{n}\cdot\mathbf{B}\mathbf{M}=\int_\Omega d^3r\, \mathbf{B}\cdot\nabla\mathbf{M}$.
Therefore the last two terms in (\ref{eq:appendix_almost_done}) cancel to leave
\begin{equation}
\mathbf{f} = \int_{\Omega} d^3r \, (\nabla\mathbf{B})\cdot\mathbf{M},
\end{equation}
consistent with (\ref{eq:force}).

Since $\mathbf{B}=\mathbf{B}^{coils}+\mathbf{B}^{plasma}$ is curl-free in $\Omega$, $(\nabla\mathbf{B})\cdot\mathbf M - \mathbf{M}\cdot\nabla\mathbf{B}=\mathbf{M}\times(\nabla\times\mathbf{B})$ vanishes in $\Omega$.
Thus, we also have $\mathbf{f} = \int_{\Omega} d^3r \, \mathbf{M}\cdot\nabla\mathbf{B}$.
%= \mu_0 \int_{\Omega} d^3r \, \mathbf{M}\cdot\nabla\mathbf{H}$, where $\mathbf{H}$ (like $\mathbf{B}$) is the field from sources outside $\Omega$.

%%       \citet{<label>} ==> Lamport [21]
%%       \citep{<label>} ==> [21]
%%
% Example citation, See \citet{hirshman1983steepest}.
% Another citation \cite{hirshman1983steepest}.

\bibliographystyle{elsarticle-num-names} 
\bibliography{Ferromagnetic_components.bib}




% %% For authoryear reference style

% \bibitem[Lamport(1994)]{lamport94}
%   \textit{\LaTeX: a document preparation system},
%   Addison Wesley, Massachusetts,

% \end{thebibliography}
\end{document}